\documentclass{article}

\input{psfig.sty}
\usepackage {graphicx}
\usepackage{amsmath} 
\usepackage{url}

\begin{document}
\pagestyle{headings}

%\mainmatter

\title{Solving NP-complete problems with delayed signals: an overview of current research directions}

\author{Mihai Oltean, Oana Muntean\\
Department of Computer Science,\\
Faculty of Mathematics and Computer Science,\\
Babe\c s-Bolyai University, Kog\u alniceanu 1,\\
Cluj-Napoca, 400084, Romania.\\
\{moltean,oanamuntean\}@cs.ubbcluj.ro\\
http://www.cs.ubbcluj.ro/$\sim$moltean/optical\\
}

\maketitle

\begin{abstract}

In this paper we summarize the existing principles for building unconventional computing devices that involve delayed signals for encoding solutions to NP-complete problems. We are interested in the following aspects: the properties of the signal, the operations performed within the devices, some components required for the physical implementation, precision required for correctly reading the solution and the decrease in the signal's strength. Six problems have been solved so far by using the above enumerated principles: Hamiltonian path, travelling salesman, bounded and unbounded subset sum, Diophantine equations and exact cover. For the hardware implementation several types of signals can be used: light, electric power, sound, electro-magnetic etc.

\end{abstract}

\textbf{keywords: }{unconventional computing, signal-based computing, NP-complete, delay lines, optical computing}

\section{Introduction}

NP-complete problems \cite{garey} have attracted a great number of researchers due to their simple terms but huge complexity. Despite the impressive amount of work invested in these problems no one has been able to design a polynomial-time algorithm for them. A relatively new direction is to attack these problems with unconventional devices. DNA computers \cite{adleman}, Quantum computers \cite{Feynman,Shor}, bubble soap \cite{aaronson}, membrane computing \cite{paun,paun2}, gear-based computer \cite{Vergis}, adiabatic algorithm \cite{Kieu} etc are few of the most important approaches of this kind.

Here we outline some of the most important principles governing some unconventional devices which use delayed signals for encoding solutions to NP-complete problems. A common feature of all these devices is the fact that the signals are delayed by a certain amount of time. The existence of a solution is determined by checking whether there is at least one signal which was delayed by a precise amount of time. If we don't find a signal at that moment it means that the problem has no solution. 

The difficulty of this approach resides in the design of a delaying system such that the solution can simply be read at an exact moment of time.

At the current stage we are interested to find only if a solution exists for the investigated problem. Otherwise stated, we try to solve the decision (YES/NO) version of the problems. 

Since we work with signals we need a physical structure in which the signals travel. The structure is usually represented as a directed graph with arcs connecting nodes. The directed graph is designed in such a way that all possible solutions of the problem are generated. The device has 2 special nodes: a start node (where the signal enters) and the destination node (where the signals are collected and interpreted).

Initially, the signal (pulse) is sent to the start node. As the signal traverses in graph it will be divided into more and more signals. Each of them will encode a partial solution for the problem. It is important that the signals do not annihilate each other. At the destination node we filter the solutions by checking for the good ones.

There are several other ways for solving NP-complete problems by using light and its properties. Two other different approaches have been presented in \cite{Collings} and \cite{Shaked}. 

The paper is organized as follows: Section \ref{NP} describes the NP-complete problems. Properties of the signal useful for our research are described in section \ref{properties}. Operations performed in our devices are described in section \ref{operations}. Some examples of devices working with delayed signals are shown in section \ref{examples}. Several practical aspects for hardware implementation are discussed in sections \ref{precision} - \ref{hard}. Difficulties for the practical implementations are given in section \ref{difficult}. Further work directions are given in section \ref{automation}. Section \ref{conclusion} concludes our paper.

\section{YES/NO NP-complete problems}
\label{NP}

NP-complete problems \cite{garey} are a special class of problems for which we don't know whether a polynomial-time algorithm exists. There is no proof that we can solve them only in exponential time nor a polynomial algorithm was proposed so far. NP-complete problems are linked together by a polynomial time reduction. Thus, if one of them is solved in polynomial time it means that all others can be solved in polynomial-time.

NP-complete problems are usually formulated as decision problems. Instead of asking for a minimal solution (e.g. the shortest path, the smallest set, the lowest point etc) one can ask if there is a solution smaller than a fixed constant $K$ (e.g. the length of the path is shorter than $K$, the number of elements in the set is smaller than $K$ etc). These are decision problems (also known as YES/NO problems).

In our research we are interested in the solving the decision problems. We are not interested to find the actual solution of the problems.

\section{Properties of the signal that we can count on}
\label{properties}

Two properties of signal are of great interest for our research. Most types of signal that we know (light, sound, electric etc) have these properties.

\begin{itemize}

\item{The speed of the signal has a limit. We can delay any signal by forcing it to pass through a cable of a certain length.}

\item{The signal can be easily divided into multiple signals of smaller intensity/power.}

\end{itemize}

For some problems it is required for the signal makes some loops (see the unbounded subset sum problem in section \ref{examples}). This type of flow is not possible for the electric-based signals. This is why, when we talk about problems whose structure requires loops we assume that we work with optical signals not with electric-based.

\section{What we do with the signals}
\label{operations}

The following manipulations of the signals are performed within the devices:

\begin{itemize}

\item{When the signal passes through an arc it is delayed by the amount of time assigned to that arc.}

\item{When the signal passes through a node it is divided into a number of signals equal to the out degree of that node. Every obtained signal is directed toward one of the nodes connected to the current node. In this way we add parallelism to our devices. This feature is actually the source for a major drawback: due to repetitive divisions the strength of the signal decreases exponentially and more and more powerful signals are required for larger and larger instances of the problems.}

\end{itemize}

\section{Basic idea}
\label{basic}

The device has a directed graph-like structure with 2 special nodes: a start node and a destination node.

The signal is sent initially to the start node. It will then enter in the rest of the graph where the actual computations are performed.

The purpose of the destination is to collect the solutions. In the destination node we have placed a reading device which measures the moments when the signals arrive there.

In the rest of the directed graph we have nodes which split the signal and arcs which delay the signal. We work with arcs (directed edges) instead of edges (undirected edges) because we don't want to annihilate the signals coming from 2 opposite directions.

\begin{figure}[htbp]
\centerline{\includegraphics[width=4.00in,height=1.52in]{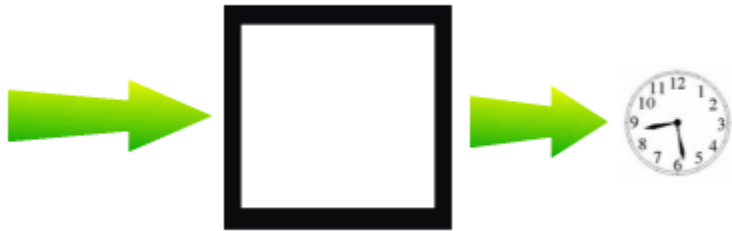}}
\label{fig_black_box}
\caption{Black-Box design of our device. Signal enters from start node. Within the device, the signals follow different paths and are divided multiple times. In the destination we will get different signals at different moments of time.}
\end{figure}

The graph must be constructed in a special way. Each signal follows a particular path meaningful for the problem structure. The signal constructs a solution by visiting nodes and arcs. When it traverses an arc it is delayed by some amount of time. Finally it arrives in the destination node. A black-box representation of our device is given in Figure \ref{fig_black_box}.

If the solution of the problem was correctly constructed we will have a particular delay induced to a particular signal. In what follows we denote by $B$ this delay. No other signals (which do not encode solutions) can have this delay. This is a hard constraint. The delaying system must obey this rule, otherwise we cannot make distinction between signals representing complete solutions and signals encoding partial or incorrect solutions.

The difficulty of this approach resides in satisfying this constraint. 

\section{Did we solve the problem?}
\label{solution}

In the destination node we have more signals arriving at different moments. There can be two cases:

\begin{itemize}

\item{If there is a signal arriving at moment $B$, this means that there is a solution for our problem.}

\item{If there is a no signal arriving at moment $B$ means that there is no solution to our problem.}

\end{itemize}

If there is more than one signal arriving at the moment $B$ in to the destination it simply means that there are multiple solutions to the problem. This is not a problem because we want to answer the YES/NO decision problem (see section \ref{NP}). At this moment we are not interested in finding the actual content of the solution.

Because we work with continuous signal we cannot expect to have discrete output at the destination node. This means that arrival of the signals is notified by fluctuations in the intensity of the signal. These fluctuations will be read by some specialized device (such as an oscilloscope).

\section{Designing the graph}
\label{examples}

\begin{figure}[htbp]
\centerline{\includegraphics[width=4.5in,height=5in]{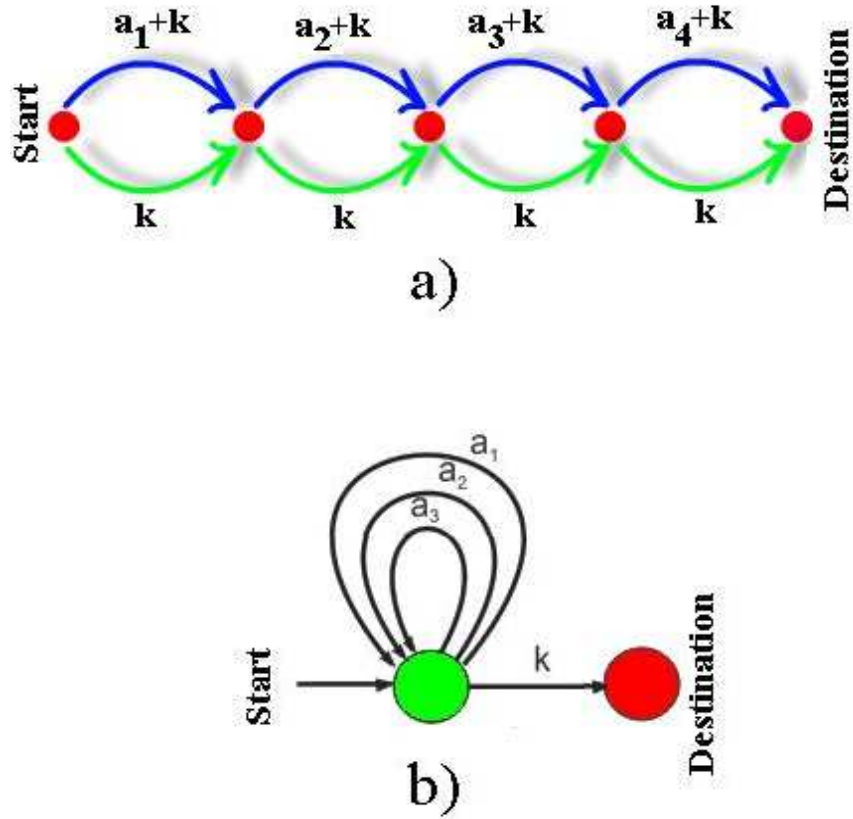}}
\caption{The graphs representing devices for a) standard subset sum, b) unbounded subset sum. A problem with 4 numbers \{$a_1, a_2, a_3, a_4$\} is considered for the standard subset sum. The constant $k$ was added to all arcs because we cannot have cables of length 0. All arcs have been depicted similarly, but in reality they can have different lengths depending on the values of the numbers in the given set. A problem with 3 numbers \{$a_1, a_2, a_3$\} is considered for the unbounded subset sum. Signals encoding combinations of numbers arrive in the internal nodes and are sent either to destination or back again for adding more delays}
\label{fig1}
\end{figure}

\begin{figure}[htbp]
\centerline{\includegraphics[width=4in,height=5.4in]{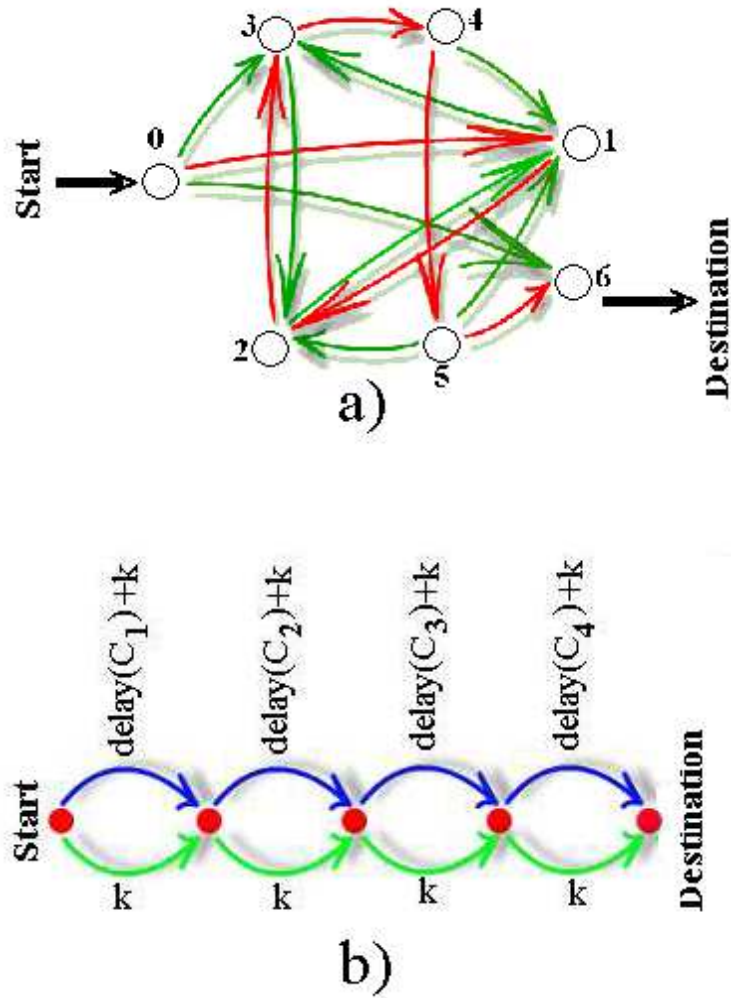}}
\caption{The graphs representing devices for a) Hamiltonian path, b) Exact Cover. In the case of Hamiltonian path, the length of the arcs connecting nodes is not important - it can be the same for all arcs. What is important are the delays induced by each node. Thus, inside each node we have another set of arcs - not depicted here - which introduce some delays. The device for Exact Cover is more complicated. Since we wanted that each number from the original set to appear exactly once we use the delaying system from the Hamiltonian path. However, this is not enough, because here we have to select sets from a given collection and not single numbers. For this purpose we have assigned to each set a delay equal to sum of delays attached to numbers from that set. It is denoted by $delay(C_i)$. Choosing the correct set of sets is done in a similar manner with the standard subset sum (see Figure \ref{fig1}). Note that numbers from original set cannot be seen in this picture. They are actually hidden inside the delays induced by each set }
\label{fig2}
\end{figure}

\begin{figure}[htbp]
\centerline{\includegraphics[width=3.5in,height=5.5in]{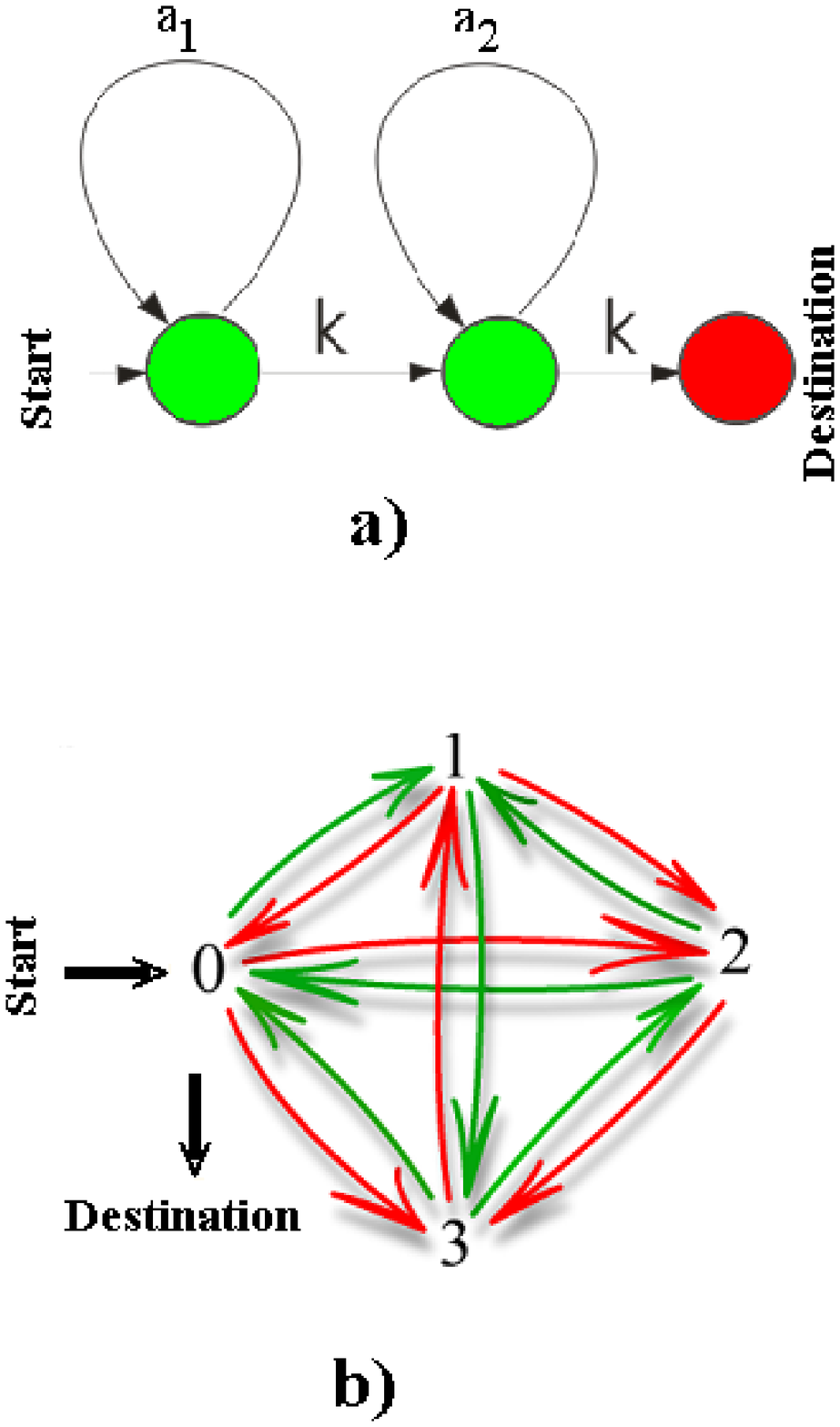}}
\caption{The graphs representing devices for: a) Diophantine equations, b) travelling salesman. A loop of the device for Diophantine equations is similar to the device for unbounded subset sum problem with only one number in the set. Signals looping through device are actually increasing values for $x$ and $y$. Note that the device can be extended for any number of variables. Travelling salesman is again a complicated device. Practically $n!$ Hamiltonian paths can be generated and we are interested in searching for the shortest one. First step is to ensure that we can distinguish between Hamiltonian and non-Hamiltonian paths. This is done as it was explained in Figure \ref{fig1}. To check for the shortest path we add some delays for each arc connecting nodes. These delays must be significantly shorter than the delays within nodes so that the discovery of Hamiltonian paths is not affected}
\label{fig3}
\end{figure}

Figures \ref{fig1}, \ref{fig2} and \ref{fig3} show several directed graphs used for solving various problems. All graphs have a polynomial number of nodes. In what follows we describe the basic ideas behind each device.

The standard subset sum problem has the simplest design \cite{Muntean,oltean_subset_sum}. Each number can appear or not in the final solution. This decision is represented in our device by 2 arcs having the same 2 nodes as extremities. One of the arcs has 0 length and the other arc delays the signal by an amount of time equal to one of the numbers from the given set. If the signal traverses the arc having the length greater than 0 it means that the corresponding number is selected in the solution. If the signal traverse the 0 length arc it means that the corresponding number is not selected in the solution. Practically we cannot have arcs of 0 length. This is why a constant $k$ is added to all arcs. ($B + n*k$) is the moment when the existence of a solution should be checked (where $B$ is the target value of the problem, $n$ is the cardinal of given set and $k$ is a constant). Finally, since we want to sum all numbers in solution we construct the device in a serializable way (see Figure \ref{fig1} a)). The delays are polynomial (in the size of the given numbers).

The design for unbounded subset sum contains fewer arcs compared to standard subset sum due to constraint relaxation (each number can appear multiple times in the solution). This is why we don't have to create a serial structure with no return arcs. Instead a loop structure was proposed \cite{Muntean,Muntean_unbound}. The internal node is used for dividing any incoming signal into $n+1$ subsignals (where $n$ is the cardinal of the given set). $n$ signals are sent back to arcs encoding numbers and the $(n+1)^{th}$ is sent to the destination (see Figure \ref{fig1} b)).

The Hamiltonian problem asks to visit each city exactly once (see Figure \ref{fig2} a)). It is not easy to satisfy this constraint since we cannot restrict the signal to visit a node exactly once. More than that, the distance between nodes is not important in this problem, thus if incorrectly designed can lead to multiple rays arriving in the same moment in destination. Because the constraints are imposed by nodes, the delays should be focused on nodes instead of arcs. This is different from the previously described solutions where the only purpose of nodes was to divide the signal. Let us suppose that the signal encoding the Hamiltonian path arrives at moment $B$ in the destination. No other signals (not encoding Hamiltonian paths) must arrive in the same moment there. We have to choose the delay induced by each node in order to satisfy this constraint. In \cite{oltean_uc,oltean_nc} it was shown an example of such delaying system.  That system guarantees that the delay induced to the signal encoding a Hamiltonian path will not be equal to any other path visiting some cities more than once or skipping some other cities. Unfortunately it was exponential (the length of the delays increases exponentially with the number of nodes).

The directed graph for the Exact Cover problem \cite{garey} (see Figure \ref{fig2} b)) is a combination between Hamiltonian path and standard subset sum \cite{oltean_exact_cover}. Some subsets from a collection must be chosen (like in the standard subset sum) and each number from the original set must appear exactly once (like nodes in the Hamiltonian path). The delaying system is exponential because it is based on the Hamiltonian path device.

For solutions to Diophantine equations we have to choose some positive integer numbers $x$ and $y$ which have the property $a_1 * x + a_2 * y = c$ \cite{Muntean,Muntean_eq} (where $a$ and $b$ are some positive integer numbers). A brute-force approach was employed by generating all possible pairs ($x$, $y$). The trick consists in a loop whose purpose is to increase the value of a variable with 1 unit. The signal enters in the loop and traverses it. When exits it will be divided into 2 signals: one of them will be sent to the next node and another one will be sent back to the loop for increasing the delay another time-unit (see Figure \ref{fig3} a)). Because we cannot have cables of length 0 we have to search for a solution at moment $c + 2*k$, where $k$ is the delayed induced by cables connecting the nodes. The delaying system is polynomial.

The construction of TSP device imposes a double difficulty: some nodes must be visited exactly once and the total path must be the shortest possible \cite{haist,haist_erratum}. If we ignore delays on nodes we will have paths not being Hamiltonian. If we focus only on delays induced to nodes, we will not obtain the shortest path. To solve this difficulty we assign a large (and exponential) delay to each node and smaller delays on arcs. Only Hamiltonian paths (visiting each node) are checked for the length of incorporated arcs.

Table \ref{tab:delay_length} shows the length of the cables used for delaying the signals. 3 problems requires exponential delays. The other 3 problems require polinomial time length for cables. However, even if we have cables of polynomial length is not enough because the energy consumption is exponential with the size of the instance.

\begin{table}[htbp]
\begin{center}
\caption{The magnitude of delays required for solving the problems investigated in this paper}
\label{tab:delay_length}
\begin{tabular}
{p{150pt}p{100pt}}
\hline
Problem& 
delays magnitude \\
\hline
Standard subset sum& 
polynomial \\
unbounded subset sum& 
polynomial \\
Hamiltonian path& 
exponential \\
Exact Cover& 
exponential \\
Diophantine equations& 
polynomial \\
Travelling salesman& 
exponential \\
\hline
\end{tabular}
\end{center}
\end{table}

\section{Precision}
\label{precision}

A problem is that we cannot measure the moment $B$ exactly. We can do this measurement only with a given precision which depends on the tools involved in the experiments.

Let us denote by $P$ the precision used for reading our signals. This means that we should not have two signals that arrive at 2 consecutive moments at a difference smaller than $P$. We cannot distinguish them if they arrive in an interval smaller than $P$. In our case, it simply means that if a signal arrives in to the destination in the interval $[B-P, B+P]$, we cannot be perfectly sure that we have a correct subset or one which does not have the  property in question.

\subsection{What if we delay by cables ?}

Let us denote by $v$ the speed of the signal. Based on that we can easily compute the minimal cable length that should be traversed by the signal in order (for the latter) to be delayed with $P$ seconds. This is obviously $v * P$ meters. This value is the minimal delay that should be introduced by an arc. Assuming a $3*10^8 m/s$ for the optical signal and a $10^{-12}$ precision of the best oscilloscope we get a $3*10^{-4}m$ for the shortest cable that we can have in our system.

This value is the minimal delay that should be introduced by an arc in order to ensure that the difference between the moments when two consecutive signals arrive at the destination node is greater than or equal to the measurable unit of $P$ seconds. This will also ensure that we will be able to correctly identify whether the signal has arrived in to the destination node at a moment equal to the sum of delays introduced by each arc. 

A constraint is that all the lengths must be integer multiples of $v*P$. We cannot accept cables whose lengths can be written as $x*v*P + y$, where $x$ is an integer and $y$ is a positive real number lower than $v*P$ because by combining this kind of numbers we can have a signal in the above mentioned interval but that signal does not encode a subset whose sum is the expected one.

Once we have the length for that minimal delay, is quite easy to compute the length of the other cables that are used in order to induce a certain delay.

Note that the maximal number of nodes can be increased when the precision of our measurement instruments is increased.

\section{Energy consumption}
\label{decrease}

Within nodes the signals are divided into (sub)signals. Because of that, the intensity of the signal decreases. In the worst case we have an exponential decrease of the intensity. For instance, the intensity of the subsignals will decrease $k$ times (compared to the incoming strength) if we divide each signal in $k$ subsignals. If we do this operation $n$ times we get signals $k^n$ times weaker than the original signal. Even if we have a small branching factor (the smallest possible is 2 - utilized in the solution for the subset sum \cite{oltean_subset_sum}) we still get a huge decrease for 100 nodes.

This means that our devices require a huge amount of energy for solving large instances of the problems. Actually, the consumed energy increases exponential with the size of the instance.

Please note that this difficulty is not specific to our system only. Other major unconventional computation paradigms, trying to solve NP-complete problems, share the same fate. For instance, as noted in \cite{Hartmanis}, a quantity of DNA equal to the mass of the Earth is required in order to solve Hamiltonian Path Problem instances of 200 cities using a DNA computer.

\section{Speed matters: slower is better}
\label{speed}

Assume again that we work with cables for delaying the signal.

The speed of the signal is an important parameter in our device. Working with a high speed signal is bad for our device due to the precision problems exposed in section \ref{precision}. We can either increase the precision of our measurement tools or decrease the speed of signal.

By reducing the speed of signal by 7 orders of magnitude, we can reduce the size of the involved cables by a similar order (assuming that the precision of the measurement tools is still the same). This can help us solve larger instances of the problem.

\section{Basic components for physical implementation}
\label{hard}

For implementing the proposed device we need the following components:

\begin{itemize}

\item{a source of signal (laser, pulse generator etc),}

\item{Several splitters for dividing a signal into multiple subsignals. If we work with electric signals the split is trivial. If we work with light we need some beam-splitters (such as half-silvered mirrors).}

\item{A device for reading the fluctuations in the signal intensity. If we work with electric signals we need an oscilloscope. If we work with optical signals we need either a combination of a photodiode and an oscilloscope or a special device for reading optical signals. Another possibility for optical signals is to use white light interferometry \cite{haist}.}

\item{A set of cables used for connecting nodes and for delaying the signals. }

\end{itemize}

\subsection{How to introduce delays ?}
\label{how_to_delay}

There are several ways in which the signals can be delayed. These variants depend on the type of signal to be delayed.

\begin{itemize}

\item{delay lines (optical or electrical). Electric delays are induced by either long lines or by discrete inductors and capacitors \cite{wiki}.}
\item{columns of mercury (for delaying sound waves). These devices have been originally used as memory in old computers.}

\end{itemize}

\section{Difficulties}
\label{difficult}

Several difficulties might be encountered during the construction of such devices. Some of them are listed below:

\begin{itemize}

\item{Building a general purpose device able to solve a wide range of problems and instances. This is a critical aspect for making these devices practical. Further issues are discussed in section \ref{automation}, }
\item{Setting the delays to an exact value. If we work with cables we have to cut them with huge precision. Electrical delay lines have a non-zero rise time which can introduce further difficulties to the system,}
\item{Providing enough power to the system in order to be able to solve large instances of the problems. This is the greatest difficulty for our device and cannot be solved unless P = NP, }
\item{Finding high precision reading instruments. Due to high speed of the signals we need very good reading instruments for detecting very small and very fast fluctuations in the intensity of the signal.}

\end{itemize}

\section{Automation}
\label{automation}

Currently the design and construction of graphs for each problem is made by hand. This dramatically reduced the area of applicability. Automating the process of building the devices would represent a huge step for practical applications. For achieving this purpose we have to use/design the followings: 

\begin{itemize}

\item{a scalable and reconfigurable graph. This should allow us to enable/disable arcs between nodes. The graph should be large enough to accommodate various sizes of the problems.}

\item{several programmable / reconfigurable delay lines. In this way we can easily modify the delay quantity induced by each arc. Electrical delay lines with up to 256 steps are already available on the market, which means that we can easily have 256 possible values for delays. By serializing such devices we can have larger ranges of values.}

\end{itemize}

\section{Conclusions}
\label{conclusion}

The way in which signal can be used for performing useful computations has been investigated in this paper. The techniques are based on 2 properties of the signals: the massive parallelism and the limited speed. 

Several important aspects have been exposed in this survey: what kind of operations are performed with the signals, how to construct the graph for several problems, how to find if the problem was solved or not, how to cope with precision and power decrease, which are the basic components required for implementation and which are the most common difficulties encountered during the physical implementations.

By using the described methods several problems have been solved so far: Hamiltonian path, travelling salesman, bounded and unbounded subset sum, Diophantine equations and exact cover.

Future works directions are focused on: implementing the presented devices, solving new problems and automating the construction process.

\end{document}